\documentclass[aps,preprint,showpacs,amsmath,amssymb,epsfig]{revtex4}
\usepackage{graphicx}
\usepackage{bm}

\begin{document}

\title{Experimental determination of the effective strong coupling constant}

\author{
A. Deur,$^{\njlab}$
V. Burkert,$^{\njlab}$
J.P. Chen,$^{\njlab}$
W. Korsch$^{\nuk}$ 
}

\affiliation{
\baselineskip 2 pt
\centerline{{$^{\njlab}$Thomas Jefferson National Accelerator Facility, Newport
News, VA 23606}}
\centerline{{$^{\nuk}$University of Kentucky, Lexington, KY 40506}}
}

\newcommand{\njlab}{1}
\newcommand{\nuk}{2}

\begin{abstract}
 
\emph{We present a first attempt to experimentally extract an effective 
strong coupling constant that we define to be a low $Q^2$ extension of 
a previous definition by S. Brodsky et al. following an initial work of 
G. Grunberg.
Using Jefferson Lab data and sum rules, we establish its $Q^2$-behavior over 
the complete $Q^2$-range. The result is compared to effective coupling 
constants inferred from different processes and to 
calculations based on Schwinger-Dyson equations, hadron spectroscopy or 
lattice QCD. Although the connection between the experimentally extracted 
effective coupling constants and the calculations is not established it is
interesting to note that their behaviors are similar.}

\end{abstract}


\maketitle

At experimentally accessible distances, the strong force remains the 
only interaction that resists satisfactory understanding. Quantum 
Chromodynamics (QCD), the gauge theory of the strong force, is well known at 
short distances ($\lesssim10^{-16}$ m) where it is solvable perturbatively. 
QCD, however, is not perturbatively calculable at larger distances, typically 
the scale of the nucleon radius. Recent precision data on moments of nucleon 
structure functions~\cite{eg1a proton,eg1a deuteron,E94010-1,E94010-2} reveal 
a smooth transition 
from small to large scales, while in contrast, a feature of perturbative QCD 
(pQCD) is that at $\Lambda_{QCD}$, the running strong coupling 
constant $\alpha_{s}$ becomes infinite. An approach using an effective 
coupling constant could reconcile these two seemingly 
paradoxical aspects of strong interaction.

In lepton scattering a scale at which the target structure is probed is 
given by the inverse of $Q^2$, the square of the four-momentum 
transfered to the target. One way to extract $\alpha_s$ at large $Q^2$ is 
to fit the $Q^2$-dependence of the moments of structure functions. Among all 
moments, the Bjorken sum rule~\cite{Bjorken} is a convenient relation for such 
an extraction~\cite{Ellis-Karliner}. Furthermore, as will be discussed, the 
Bjorken sum may offer unique advantages to define a effective coupling at 
low $Q^2$.

In the limit where the energy transfer $\nu$ and $Q^{2}$ are infinite, while 
$x\equiv Q^{2}/(2M \nu)$ remains finite ($M$ is the nucleon mass), the 
Bjorken sum rule reads:
\begin{equation}
\Gamma_{1}^{p-n}\equiv\Gamma_{1}^{p}-\Gamma_{1}^{n}\equiv\int_
{0}^{1}dx(g_{1}^{p}(x)-g_{1}^{n}(x))=\frac{1}{6}g_{A}.
\end{equation}
\noindent 
$g_{1}^{p}$ and $g_1^{n}$ are spin structure functions for the proton and 
neutron. The axial charge of the nucleon, $g_{A}$, is known from neutron 
$\beta$-decay. for finite $Q^2$ much greater than $\Lambda_{QCD}^{2}$, 
the $Q^2$-dependence of the Bjorken sum rule is given by a double series in 
$Q^{2-t}$ ($t$=2,4... being the twist) and in $(\alpha_{s}/\pi)^{n}$. The 
$(\alpha_{s}/\pi)^{n}$  series is given by pQCD evolution equations. At 
leading twist ($t=2$) and $3^{rd}$ order 
in $\alpha_{s}$, in the $\overline{\textrm{MS}}$ scheme and dimensional 
regularization, we have~\cite{Larin}:
\begin{eqnarray}
\Gamma_{1}^{p-n}=\frac{1}{6}g_{A} [ 1-\frac{\alpha_{s}}{\pi}-3.58
\left(\frac{\alpha_{s}}{\pi}\right)^{2}
-20.21\left(\frac{\alpha_{s}}{\pi}\right)^{3} ]. 
\label{eqn:bj}
\end{eqnarray}
\noindent 
The validity of the sum rule is verified at $Q^{2}$=5 GeV$^2$ to better than 
10\%~\cite{SLAC}.

The extraction of the Bjorken integral using data from the Thomas Jefferson 
National Accelerator Facility (JLab) in the $Q^{2}$-range of 0.17-1.10 
GeV$^{2}$ has been reported recently~\cite{JLab}. The use of Eq.~\ref{eqn:bj} 
as an Ansatz for definition of an effective running coupling
constant at low $Q^2$ has ambiguities and difficulties as 
well as some practical advantages. The advantages are the following: 
firstly, it was pointed out in Ref.~\cite{Altarelli} that the extraction 
of $\alpha_{s}$ does not depend strongly on the low-x extrapolation.
Secondly, this flavor non-singlet contribution does not mix quark and 
gluon operators when evolved. Hence the pQCD evolution is known to higher 
order.
Thirdly, the JLab data are at constant $Q^{2}$. This avoids 
a possible ambiguity encountered in previous experiments, namely that 
in order to combine neutron and proton data, structure functions must be 
evolved to a common $Q^{2}_0$, which needs  $\alpha_{s}$ itself as input.
This is especially important in our case since we cannot anticipate the value
of  $\alpha_{s}$ at such low $Q^2$, due to the breakdown of pQCD. Hence,
no evolution of the data to $Q^2_0$ is possible. 
Difficulties within the pQCD approach are: firstly, at low $Q^{2}$ higher
twist effects become important and are not well known~\cite{JLab}. Secondly, 
the pQCD expansion loses its meaning as $Q$ approaches $\Lambda_{QCD}$ where, 
as a consequence of renormalization and regularization, 
$\alpha_{s}^{pQCD}$ itself is singular. It is
necessary to use an appropriate theoretical framework to circumvent these
difficulties. Such frameworks have been developed, see for 
example~\cite{brodsky} and~\cite{Dokshitzer}. We use the method of 
``effective charges'' of Grunberg~\cite{grunberg 1},  where the 
non-perturbative terms and higher order perturbative processes are absorbed in 
the definition of the coupling constant. In our case, it obeys the following 
definition:
\begin{eqnarray}
\Gamma_{1}^{p-n}\equiv \frac{1}{6}g_{A} [ 1-\frac{\alpha_{s,g_1}}{\pi} ]. 
\label{eqn:bjltlo}
\end{eqnarray}
\noindent 
Eq.~3 provides a definition of an effective QCD running coupling
that we will explore here. The inherent systematic uncertainties in this 
experimental Ansatz, and in those of the various theoretical approaches are 
unknown. Their comparison provides a framework for further analysis. 

The coupling constant defined with Eq.~3 still obeys the renormalization group 
equation $d\alpha_s(k)/dln(k)=\beta(\alpha_s(k))$~\cite{grunberg 1}. The first 
two terms in the Eq.~2 and 3 series are independent of the choice of gauge and 
renormalization scheme. Consequently, the effective coupling constant is 
renormalization scheme and gauge independent, but becomes process-dependent. 
These process-dependent coupling constants can be related by using 
``commensurate scale relations'' which connect observables without 
renormalization scheme or scale ambiguity~\cite{brodsky 1}. In this topic, 
an important relation is the Crewther relation~\cite{crewther,brodsky 1}.
 
Considering an effective coupling constant yields other advantages beside 
renormalization scheme/gauge independence: such a procedure improves
perturbative expansions~\cite{grunberg 1,brodsky 1}, the effective charge
is analytic when crossing quark mass thresholds, is non-singular at 
$Q=\Lambda_{QCD}$, and is well defined at low $Q^{2}$~\cite{brodsky 2}.
The choice of defining the effective charge with Eq.~3 has unique advantages: 
Low $Q^2$ data exist and near real photon data will be available soon from 
JLab~\cite{gdh neutron,gdh proton}. Furthermore, sum rules constrain
$\alpha_{s,g_1}$ at both low and high $Q^2$ limits, as will be discussed in
the next paragraph. Another advantage is that $\Gamma _1^{p-n}$ is a quantity 
well suited to be calculated at any $Q^2$ because of various cancellations
that simplify calculations~\cite{Volker,JLab}. 

\begin{figure}[ht!]
\begin{center}
\centerline{\includegraphics[scale=0.7, angle=0]{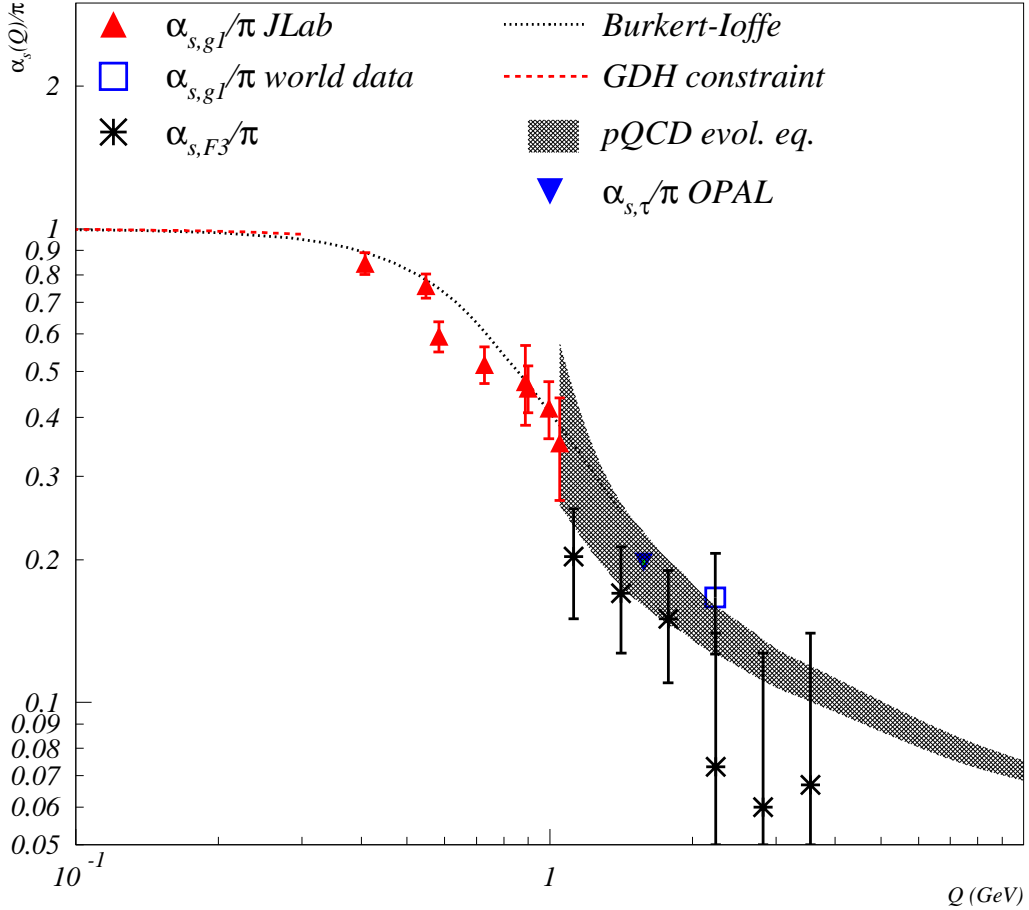}}
\end{center}
\vspace{-1cm}
\caption{(color online) $\alpha_{s}(Q)/\pi$ obtained from JLab data (up triangles), 
the GLS sum result from the CCFR collaboration~\cite{CCFR} (stars), the world 
$\Gamma^{p-n}_1$ data (open square), the Bjorken sum rule (gray band) and 
the Burkert-Ioffe Model. $\alpha_{s}(Q)/\pi$ derived using leading order
commensurate scale relations and the $\alpha_{s, \tau}(Q)/\pi$ from 
OPAL data is given by the reversed triangle. The dashed line is the GDH 
constraint on the derivative of $\alpha_{s,g_1}/ \pi$ at $Q^2$=0.}
\label{fig:alpha}
\end{figure}

Using Eq.~\ref{eqn:bjltlo} and the JLab data, $\alpha_{s,g_1}$ can be 
formed. The elastic ($x=1$) contribution is excluded in 
$\Gamma_{1}^{p-n}$. The resulting $\alpha_{s,g_1}/\pi$ is shown in 
Fig.~\ref{fig:alpha}. Systematic effects dominate the uncertainties, see 
ref.~\cite{JLab} for details. The uncertainty from $g_A$ is small. We also 
used the world data of the Bjorken sum evaluated at $<Q^2>$=5 GeV$^2$ to 
compute $\alpha_{s,g_1}/\pi$. We can also use a model for $\Gamma _1$ and, 
using Eq.~\ref{eqn:bjltlo}, form $\alpha_{s,g_1}$. We chose 
the Burkert-Ioffe model~\cite{Burkert} because of its good match with the 
experimental data on moments of spin structure 
functions~\cite{eg1a proton}-\cite{JLab}. It is 
interesting to note the behavior of $\alpha_{s,g_1}$ near $Q^{2}=0$ where it 
is constrained by the Gerasimov-Drell-Hearn (GDH) sum rule~\cite{gdh} that 
predicts the derivative of the Bjorken integral with respect to $Q^{2}$ at $Q^{2}=0$,~\cite{gdh*}:
\begin{eqnarray}
\Gamma_1^{p-n}= \frac{Q^2}{16 \pi^2 \alpha}(GDH^p-GDH^n)=\frac{-Q^2}{8}
(\frac{\kappa ^2_p}{M_p^2} - \frac{\kappa ^2_n}{M_n^2})
\label{eqn:gdhcons}
\end{eqnarray}
\noindent  
where $\kappa _p$ ($\kappa _n$) is the proton (neutron) anomalous magnetic 
moment and $\alpha$ is the QED coupling constant. This, in combination with 
Eq.~3, yields:
\begin{eqnarray}
\frac{d \alpha_{s,g_1}} {dQ^2}= -\frac{3 \pi}{4g_A} \times 
(\frac{\kappa ^2_n}{M_n^2} - \frac{\kappa ^2_p}{M_p^2})
\label{eqn:gdhcons2}
\end{eqnarray}
\noindent  
The constraint is shown by the dashed line. At $Q^{2}=0$ the Bjorken sum is 
zero and $\alpha_{s,g_1}=\pi$, a particular property of the definition of 
$\alpha_{s,g_1}$. At larger $Q^2$, where higher twist effects are 
negligible, $\alpha_{s,g_1}$ can be evaluated by estimating the right hand side 
of Eq.~2 (using $\alpha_{s}^{pQCD}$ as predicted by pQCD) and equating it to 
$g_{A}[ 1-\alpha_{s,g_1}/\pi ]/6$. The resulting $\alpha_{s,g_1}$ is shown by 
the gray band~\cite{noteRS}. The width of the band is due to the uncertainty 
in $\Lambda_{QCD}$. Finally $\alpha_{s,F_3}$ can be computed using 
data on the Gross-Llewellyn Smith sum rule \cite{GLS}, which relates the 
number of valence quarks in the hadron, $n_v$, to the structure function 
$F_{3}(Q^{2},x)$ measured in neutrino scattering. At leading twist, the GLS 
sum rule reads:
\begin{eqnarray}
\int_{0}^{1}F_{3}(Q^{2},x)dx=n_{v}[1-\frac{\alpha_{\rm{s}}(Q^{2})}
{\pi}-3.58\left(\frac{\alpha_{\rm{s}}(Q^{2})}{\pi}\right)^{2}
-20.21\left(\frac{\alpha_{\rm{s}}(Q^{2})}{\pi}\right)^{3} +...]. \nonumber
\end{eqnarray}
\label{eqn:GLS}
\noindent
Using the data taken by the CCFR collaboration~\cite{CCFR}, we can apply the 
same procedure as for the Bjorken sum rule to compute $\alpha_{s, \rm{F_3}}$. 
We expect $\alpha_{s, \rm{F_3}}=\alpha_{s, \rm{g_1}}$ at large $Q^2$,
since the leading twist $Q^2$-dependence of Eq.~2 and Eq.~6 is identical
up to $\alpha_{s,}^3$, up to a very small difference at order $\alpha_{s,}^3$
coming from the light-by-light contribution to the GLS sum.

In principle, the commensurate scale relations derived by Brodsky and Lu
in ref.~\cite{brodsky 1} should be applied when comparing $\alpha_{s,
\rm{g_1}}$ to $\alpha_{s,F_3}$ (and $\alpha_{s, \tau}$ in Fig.~1). In
practice, however, the resulting corrections are small; the same scale is
used in the $\alpha_{s, \rm{g_1}}$ and $\alpha_{s,F_3}$ comparison and the
correction from the light by light contribution decreases the value of
$\alpha_{s,F_3}$ by at most 1\%. These corrections were neglected. The 
commensurate scale relations should be used when comparing $\alpha_{s,\tau}(Q)$
extracted from OPAL data on $\tau$-decay~\cite{brodsky 2} to 
$\alpha_{s, \rm{g_1}}$. At leading order (the only order available since only 
one Q value of $\alpha_{s,\tau}(Q)$ is available) the Q correction necessary 
to compare $\alpha_{s,\rm{g_1}}$ to $\alpha_{s,\tau}$ leads to a $20\%$ change 
on the effective charge. Such Q correction was applied.

In summary, assuming the validity of the GDH and Bjorken sum rules for 
low and large $Q^2$ respectively and using the JLab data at intermediate 
$Q^2$, we can evaluate $\alpha_{s,\rm{g_1}}$ at any value of $Q^2$. 
The absence of divergence in $\alpha_{s, \rm{g_1}}$ is obvious since it is 
defined from finite experimental data. It is interesting to notice
that $\alpha_{s,\rm{g_1}}$ loses its $Q^2$-dependence at low $Q^2$. This 
feature was suggested by the work of Brodsky \emph{et  al.} based on 
$\tau$-decay data~\cite{brodsky 2} and by other theoretical works, as 
will be discussed below.

Many theoretical or phenomenological studies of $\alpha_s$ at low-$Q^2$ are 
available. See~\cite{Aguilar,Bloch} for reviews. The theoretical studies 
comprise Schwinger-Dyson Equations (SDE), lattice QCD, non perturbative QCD 
vacuum, and analyticity arguments. Phenomenological studies are based on quark 
model spectroscopy, low-$x$-low-$Q^2$ reactions, parametrization of nucleon 
and pion form factors, heavy quarkonia decay, and ratio of hadron production 
cross section to $\mu ^+ \mu ^-$ production cross section in $e^+ e^-$ 
annihilation.

Similarly to experimental effective charges, different 
definitions of the strong coupling constant at low $Q^2$~\cite{Tandy1} are 
possible in the theoretical calculations. How they are related is not fully 
known. Furthermore, these calculations should be viewed as indications of the 
behavior of $\alpha_s$ rather than strict predictions. Although some 
theoretical uncertainties due to parameterizations are shown, 
the existence of unknown systematic/model uncertainties should be borne in 
mind. However, it is interesting to compare the various calculations to our 
result to see whether they show common features.

\begin{figure}[ht!]
\begin{center}
\centerline{\includegraphics[scale=0.7, angle=0]{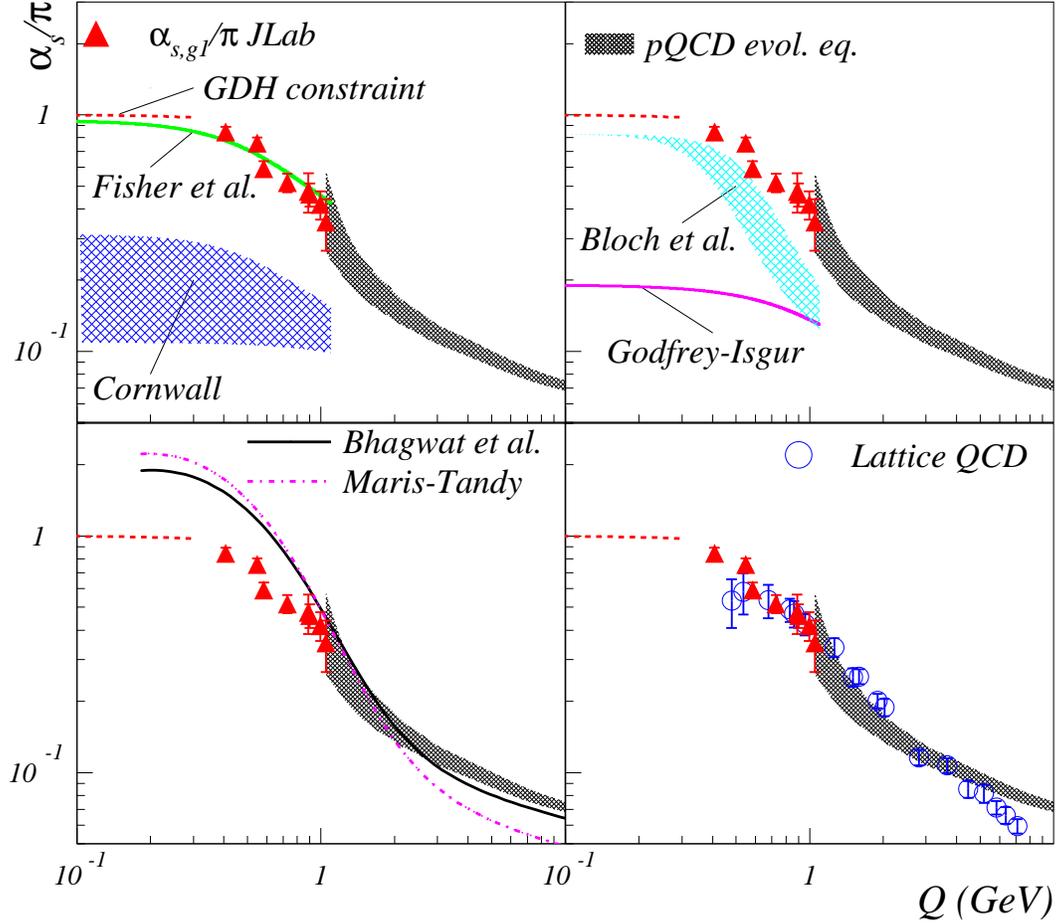}}
\end{center}
\vspace{-1cm}
\caption{(color online) $\alpha_{s,g_1}$ from JLab data and sum rules compared 
to various calculations: top left panel: SDE calculations from 
Fisher~\emph{et al.} and Cornwall; top right panel: Bloch~\emph{et al.} (SDE) 
and Godfrey-Isgur (quark model);  bottom left: Maris-Tandy (SDE) and 
Bhagwat~\emph{et al.} (SDE); bottom right: Furui and Nakajima (lattice QCD).}
\label{fig:alpha2}
\end{figure}

In theory, solving the SDE equations can yield $\alpha_s$. In 
practice, approximations are necessary and choices of approximation lead to
different values of $\alpha_s$. In the top panels and the bottom left
panel of Fig.~\ref{fig:alpha2}, we  compare our data to different
approaches~\cite{Cornwall,Fisher,Bloch2,Tandy2,Bhagwat}. The uncertainty in 
Cornwall's result is due to the uncertainties in parameters that enter
the calculation. The uncertainty on the Bloch curve is due to $\Lambda_{QCD}$. 
There is a good agreement between the absolute value obtained from the present 
Ansatz and the results of Bloch~\cite{Bloch2} and 
Fisher~\emph{et al.}~\cite{Fisher}, while the results from 
Maris-Tandy~\cite{Tandy2} and Bhagwat~\emph{et al.}~\cite{Bhagwat} do not
agree as well. The older calculation from Cornwall~\cite{Cornwall} disagree 
with the result of our Ansatz.
The Godfrey and Isgur curve in the top right panel of Fig.~\ref{fig:alpha2} 
represents the coupling constant used in a quark model~\cite{Godfrey-Isgur}.

It is interesting to notice that the $Q^2$-dependence of $\alpha_{s,g_1}$ and 
the ones of the calculations are similar. A relative comparison reveals that 
the $Q^2$-dependence of the Godfrey-Isgur, Cornwall and Fisher~\emph{et al.} 
results agree well with the data while the curves from Maris-Tandy, 
Bloch~\emph{et al.} and Bhagwat~\emph{et al.} are slightly below the data 
(by typically one sigma) for $Q^2 > 0.6$ GeV.

One Ansatz that has received recent theoretical attention for the QCD coupling 
at low $Q^2$ is related to the product of the gluon 
propagator dressing function with the ghost propagator dressing function 
squared~\cite{Fisher}. Gluon and ghost 
propagators have been computed in lattice QCD by many groups. The lattice 
results offer a fairly consistent picture~\cite{Bloch} and agree reasonably 
with the various SDE propagator results~\cite{Aguilar2}. It is calculated in 
particular in refs.~\cite{Bloch,Furui2}. In the bottom right panel of 
Fig.~\ref{fig:alpha2}, we compare  
$\alpha_{s,g_1}$ to the lattice result from ref.~\cite{Furui2}. We 
plot the lattice result that is believed to be closer to the continuum 
limit~\cite{Furui3}. There is a good 
agreement between the lattice QCD calculations and our data.

To conclude, we have formed an effective strong coupling constant 
$\alpha_{s,g_1}$ at low $Q^2$. Data, together with sum rules, allow to 
obtain $\alpha_{s,g_1}$ at any $Q^2$. The connection between the Bjorken and 
the GDH sum rules yields a value of $\alpha_{s,g_1}$
equal to $\pi$ at $Q^2$ = 0. An important feature of $\alpha_{s,g_1}$ is its 
loss of $Q^2$-dependence at low $Q^2$. We compared our result to other 
coupling constants from different reactions. They agree with each other, 
although they were defined from different processes. We also compared 
$\alpha_{s,g_1}$ to SDE calculations, lattice QCD calculations and a 
coupling constant used in a quark model. Although the relation between the 
various calculations is not well understood, the data and calculations agree 
in most cases especially when only considering $Q^2$-dependences. 
It will be interesting in the 
future  to pursue the same analysis with lower $Q^{2}$ data that will be 
available both for the neutron~\cite{gdh neutron} and 
proton~\cite{gdh proton}. 

We are grateful to S. Brodsky, S. Gardner. W. Melnitchouk and P. Tandy for 
helpful discussions. We thanks S. Furui and P.O. Bowman for sending us their
lattice results. 
This work was supported by the U.S. Department of Energy 
(DOE). The Jefferson Science Associates (JSA) operates the 
Thomas Jefferson National Accelerator Facility for the DOE under contract 
DE-AC05-84ER40150.


\begin{references}
\bibitem{noteRS} When forming $\alpha_{s,g_1}$ using the Bjorken 
sum rule (gray band), we obtain a renormalization scheme dependent quantity, 
which is in contradiction with our goal. In practice, however, at large 
$Q^{2}$ and for high enough order in $\alpha_{s}^{pQCD}$, the renormalization 
scheme dependence is not significant, which is why, for example, we can 
compare meaningfully a measured (renormalization scheme independent) Bjorken 
sum to the (renormalization scheme dependent) pQCD prediction.
\end{references}
\end{document}